\documentclass[
 reprint,
superscriptaddress,
amsmath,
amssymb,
aps,
prx,
aip,
rsi, floatfix]{revtex4-1}
\usepackage{color}
\usepackage{graphicx}% Include figure files
\usepackage{dcolumn}% Align table columns on decimal point
\usepackage{bm}% bold math
\usepackage{braket}% braket quantum symbols
\usepackage{float}% fix figure position
\usepackage{multirow}
\usepackage{hhline}
\usepackage{mathtools}
\usepackage{qcircuit}
\usepackage{lipsum}
\usepackage{hyperref}
\usepackage{xcolor}
\definecolor{dark-red}{rgb}{0.4,0.15,0.15}
\definecolor{dark-blue}{rgb}{0.15,0.15,0.4}
\definecolor{medium-blue}{rgb}{0,0,0.5}
\usepackage[all]{hypcap}%fix problems with figures and tables of the 'hyperref' package
\hypersetup{
    colorlinks, linkcolor={dark-red},
    citecolor={dark-blue}, urlcolor={medium-blue}
} %These setups replace the ugly and distracting red and green boxes of the 'hyperref' package with colored texts.
\usepackage{todonotes}

\newcommand{\nsp}{\hspace{-0.4pt}}
\newcommand{\xssp}{\hspace{0.4pt}}

\newcommand{\proj}[1]{\ket{#1}\nsp \bra{#1}}

\newcommand{\dif}{d}

\DeclareMathOperator{\tr}{tr}

\newcommand{\VV}{V}

\newcommand{\CZ}{\mbox{\small $\mathrm{CZ}$}}

\newcommand{\iSWAP}{\mbox{\small $\mathrm{iSWAP}$}}
\newcommand{\SWAP}{\mbox{\small $\mathrm{SWAP}$}}
\newcommand{\CPhase}{\mbox{\small $\mathrm{CPhase}$}}

\newcommand{\fSim}{\mbox{\small $\mathrm{fSim}$}}
\newcommand{\nn}{n}

\newcommand{\KK}{K}
\newcommand{\dd}{{d}}

\newcommand{\xx}{{x}}

\newcommand{\Hadamard}{\mbox{\small $\mathrm{H}$}}
\newcommand{\X}{\mbox{\small $\mathrm{X}$}}

\newcommand{\pdepol}{{p_\mathrm{depol}}}
\newcommand{\espam}{{\epsilon_\mathrm{spam}}}
\newcommand{\eincoh}{{\epsilon_\mathrm{incoh}}}
\newcommand{\ecoh}{{\epsilon_\mathrm{coh}}}
\newcommand{\dtheta}{{\Delta\theta}}
\newcommand{\dgamma}{{\Delta\gamma}}
\newcommand{\dphi}{{\Delta\phi}}

\usepackage{cleveref}
\crefname{equation}{Eq.}{Eqs.}
\Crefname{equation}{Equation}{Equations}
\crefname{figure}{Fig.}{Figs.}
\Crefname{figure}{Figure}{Figures}
\crefname{section}{Sec.}{Secs.}
\Crefname{section}{Section}{Sections}
\crefname{appendix}{App.}{Apps.}
\Crefname{appendix}{Appendix}{Apps.}
\crefname{paragraph}{Sec.}{Secs.}
\crefname{table}{Table}{Tables}

\begin{document}

\title{Context Aware Fidelity Estimation}
\author{Dripto M. Debroy}
\email{dripto@google.com}
\thanks{Contributed Equally}
\affiliation{Google Quantum AI, Venice, CA 90291}

\author{\'Elie Genois}
\email{elie.genois@usherbrooke.ca}
\thanks{Contributed Equally}
\affiliation{Google Quantum AI, Venice, CA 90291}
\affiliation{Institut quantique \& D\'epartement de Physique, Universit\'e de Sherbrooke, Qu\'ebec J1K 2R1, Canada}

\author{Jonathan A. Gross}
\email{jarthurgross@google.com}
\thanks{Contributed Equally}
\affiliation{Google Quantum AI, Venice, CA 90291}

\author{Wojciech Mruczkiewicz}
\affiliation{Google Quantum AI, Venice, CA 90291}

\author{Kenny Lee}
\affiliation{Google Quantum AI, Venice, CA 90291}

\author{Sabrina Hong}
\affiliation{Google Quantum AI, Venice, CA 90291}

\author{Zijun Chen}
\affiliation{Google Quantum AI, Venice, CA 90291}

\author{Vadim Smelyanskiy}
\affiliation{Google Quantum AI, Venice, CA 90291}

\author{Zhang Jiang}
\affiliation{Google Quantum AI, Venice, CA 90291}

\begin{abstract}
We present Context Aware Fidelity Estimation (CAFE), a framework for benchmarking quantum operations that offers several practical advantages over existing methods such as Randomized Benchmarking (RB) and Cross-Entropy Benchmarking (XEB). 
In CAFE, a gate or a subcircuit from some target experiment is repeated $n$ times before being measured. By using a subcircuit, we account for effects from spatial and temporal circuit context. Since coherent errors accumulate quadratically while incoherent errors grow linearly, we can separate them by fitting the measured fidelity as a function of $n$. One can additionally interleave the subcircuit with dynamical decoupling sequences to remove certain coherent error sources from the characterization when desired. 
We have used CAFE to experimentally validate our single- and two-qubit unitary characterizations by measuring fidelity against estimated unitaries.
In numerical simulations, we find CAFE produces fidelity estimates at least as accurate as Interleaved RB while using significantly fewer resources.
We also introduce a compact formulation for preparing an arbitrary two-qubit state with a single entangling operation, and use it to present a concrete example using CAFE to study $\CZ$ gates in parallel on a Sycamore processor.
\end{abstract}

\maketitle

\section{Introduction}\label{sec:intro}
Reliably understanding the structure of noise in quantum processors is vital for advancing quantum computation.
There are a variety of characterization methods available to quantify and validate the performance of individual quantum operations like state preparation, single- and two-qubit gates, measurement, and reset~\cite{eisert2020qcvv}.
Amongst these techniques, some use randomness to estimate gate fidelities efficiently, such as randomized benchmarking (RB)~\cite{knill2008randomized, magesan2011scalable, wallman2014randomized, sheldon2016characterizing, proctor2022scalable, polloreno_theory_2023}, cross-entropy benchmarking (XEB)~\cite{boixo2018characterizing, arute2019quantum}, channel spectrum benchmarking (CSB)~\cite{gu2023benchmarking}, and direct fidelity estimation (DFE)~\cite{da_silva_practical_2011, flammia_direct_2011, elben_randomized_2022}, while others use complete sets of input states, such as unitary tomography (UT)~\cite{nielsen2002quantum, baldwin2014quantum}, quantum process tomography (QPT)~\cite{d2003quantum, merkel2013self}, and gate set tomography (GST)~\cite{greenbaum2015introduction, nielsen2021gate}.
These methods all have upsides and downsides in terms of speed, scalability, and the amount of information provided.

An important problem that many existing techniques face is that as quantum computers scale, complex inter-component interactions arise, such as control crosstalk or temporal correlations caused by residual pulse tails. 
For example, experiments may use amplifiers with temperature dependent gain profiles, which could be linked to pulse duty cycle and cause circuit-dependent errors.
These effects can make the performance of a quantum operation highly context dependent~\cite{kelly2018physical, arute2019quantum, rudinger2019probing, wu2021strong}, and there is a growing need for characterization methods which account for them.

In this paper, we describe a characterization method called \emph{Context Aware Fidelity Estimation} (CAFE) which measures the fidelity between an experimentally implemented quantum operation and a reference unitary. In CAFE, we repeat a gate or a subcircuit from a target experiment $\nn$ times and measure the average gate fidelity against the reference unitary raised to the $\nn^\text{th}$ power. For example, the subcircuit can be a stabilizer extraction circuit in a quantum error correction experiment, as shown in~\cref{app: surface code}. This procedure allows us to capture context-related errors which isolated characterizations do not capture~\cite{murali2020software, sarovar2020detecting}. Since coherent errors accumulate quadratically as a function of $n$ while incoherent errors grow linearly, we can separate their contributions to the infidelity. This is in contrast with RB and XEB, where random compiling is used to twirl coherent errors into incoherent ones. 
\begin{figure*}[t!]
    \centering
    \includegraphics[width=0.88\linewidth]{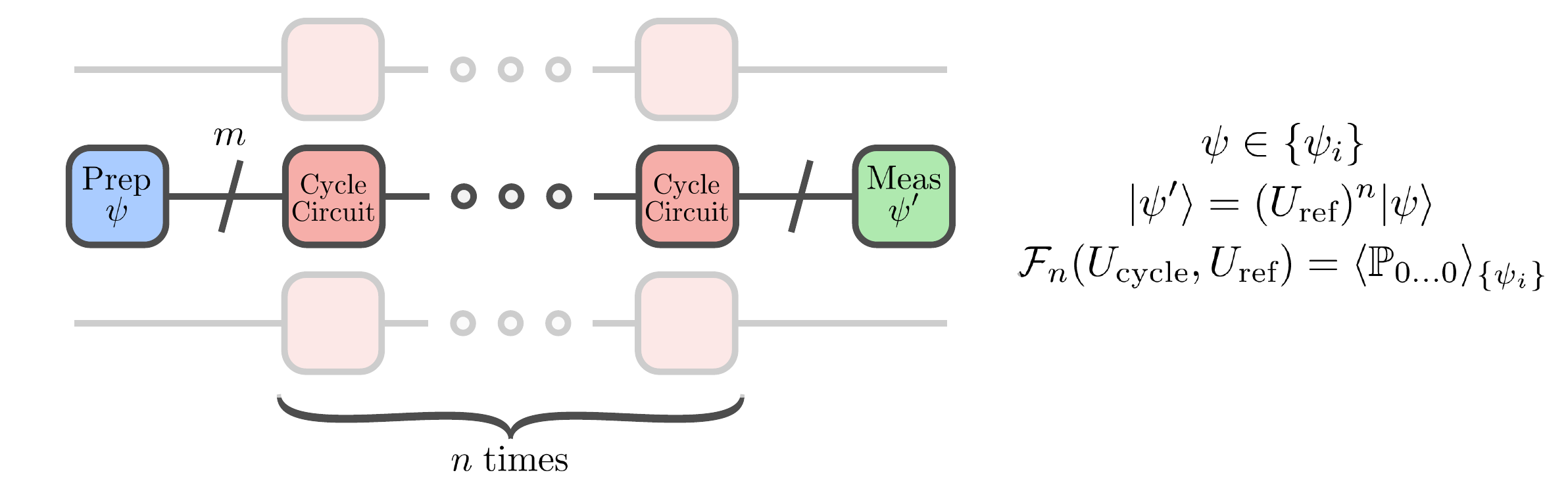}
    \caption{A schematic of the circuits used to measure the fidelity of $\nn$ repetitions of the cycle circuit, which is selected to be a subcircuit from some target experiment. First, a state pulled from a $m$ qubit 2-design, $\{\psi_i\}$, is prepared (blue). Second, the cycle circuit being characterized is applied the desired number of times $\nn$ (pink). Third, we apply the inverse of the combined preparation circuit and reference unitary $U_\mathrm{ref}$ to the $\nn^\text{th}$ power.
    Finally, we measure the resulting state in the computational basis (green). The fidelity between $n$ repetitions of the applied operation and the $n^\text{th}$ power of the reference unitary can be found by averaging the probability of getting $|0\rangle^{\otimes m}$ over all $4^m$ initial states. The faded operations represent the spatial context of the cycle circuit being characterized.
    }
    \label{fig:schematic}
\end{figure*}

Throughout the main text of this paper, we focus on using CAFE to study the performance of $\CZ$ gates implemented in parallel, as two-qubit operations have been shown to be a dominant error source across recent large-scale experiments on a number of experimental platforms, especially in the presence of stray interactions~\cite{krinner2022realizing, https://doi.org/10.48550/arxiv.2207.06431, PhysRevX.11.041058}.
In the Appendices, we present results from CAFE experiments characterizing parallel single-qubit operations as well as a circuit layer of a surface code stabilizer extraction experiment.

\section{Context Aware Fidelity Estimation}\label{sec:cafe}
A typical CAFE experiment is performed in three steps, schematically shown in \cref{fig:schematic}.
First, a state $|\psi\rangle$ is prepared from an $m$-qubit 2-design $\{\psi_i\}$~\cite{dankert2009exact, harrow2009random, zhu2010two, lu_experimental_2015}. These ensembles match the Haar-random distribution up to second moments, allowing us to measure fidelity. In App.~\ref{sec:2q prep} we present a method to construct shallow circuits that prepare arbitrary two-qubit entangled states for this purpose.
Next, the $m$-qubit circuit of interest, referred to as the \textit{cycle circuit}, is repeated $n$ times, along with operations on neighboring uninvolved qubits. The cycle circuit can include multiple operations, allowing close matching to the circuit context found in the target experiment.
As an example, an experiment which includes dynamical decoupling (DD) would be robust to certain coherent errors, and including these DD gates in the cycle circuit allows CAFE to neglect contributions from these coherent errors, focusing on the errors which impact performance.
Inserting DD gates is one of several ways for CAFE to separate coherent and incoherent contributions to gate errors, which we discuss further in Sec.~\ref{sec:coherent incoherent}.
Finally, using the ``reference unitary", we apply a circuit which ideally maps the state back to $|0\rangle^{\otimes m}$, before measuring the qubits in the computational basis. 
This final step allows us to validate the performance of different unitary characterization methods, as discussed further in Sec.~\ref{sec:comparing unitaries}. The average gate fidelity of the operation, which we refer to as \emph{fidelity} in the following, can be found by averaging over the experiments for all 2-design states $\{\psi_i\}$~\cite{nielsen2002quantum}:
\begin{equation}
    \mathcal{F} = \langle \mathbb{P}_{0...0} \rangle_{\{\psi_i\}}.
\end{equation}
We note that in the $m=2$ case, the preparation and measurement stages only use a single entangling gate, so their contribution to the total error rate is far smaller than the circuit being characterized for most values of $n$. In general, since the information we extract from CAFE comes from fitting fidelity over different values of $n$, reasonable SPAM errors do not significantly impact the characterization results, as they only cause a constant offset in the fidelity curve. While the number of 2-design states scales exponentially with qubit count, random circuits can approximate the states in polynomial depth~\cite{harrow2009random}, which could be used for performing CAFE on $m>2$ qubits.

\section{Comparing CAFE with Randomized Benchmarking}\label{sec:app-IRB-CAFE-simus}
\begin{figure}[b!]
    \centering
    \includegraphics[width=0.7\columnwidth]{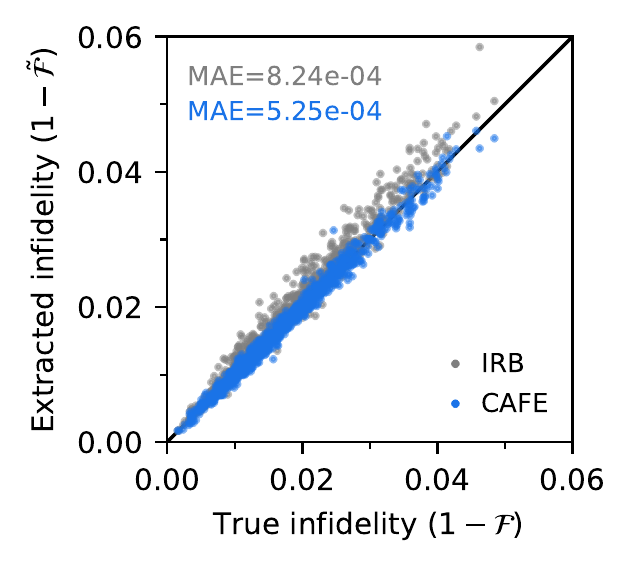}
    \caption{Simulations comparing the CAFE and Interleaved Randomized Benchmarking (IRB) techniques for characterizing the average gate infidelity of noisy $\CZ$ gates with amplitude and phase damping noise in addition to coherent errors.
    For CAFE (blue), we use depths $n \in [0,2,4,6,8]$, whereas for IRB (gray) we use depths $n \in [5,10,15,20,25,30,35]$ for obtaining both the reference RB curve and the acquisition with interleaved $\CZ$ gates.
    Both approaches use $2000$ shots per circuit.
    Labels presents the median absolute errors (MAE) of IRB and CAFE over $N=1000$ different noisy $\CZ$ gates.
    $\mathrm{MAE} = \operatorname{median}(|\Tilde{\mathcal{F}}_1 - \mathcal{F}_1|, \ldots, |\Tilde{\mathcal{F}}_N - \mathcal{F}_N|)$. We note that MAE scales with the true error of the gates being considered. 
    }
    \label{fig:app simu irb cafe}
\end{figure}
In this section, we compare our CAFE approach to the widely used Interleaved Randomized Benchmarking (IRB) protocol for estimating the fidelity of noisy $\CZ$ gates.
The error model for these gates, which is fully described in~\cref{sec:app simulations}, contains coherent errors together with amplitude and phase damping as incoherent noise.
In~\cref{fig:app simu irb cafe}, we show that for this model, and using realistic error rates, CAFE yields a more accurate average gate fidelity estimation than IRB, while requiring significantly less experimental resources.
The randomized benchmarking protocol uses 20 different circuits for 7 depths $5\le n\le35$ of random two-qubit Cliffords (which often require 2 $\CZ$s together with single-qubit gates once compiled), in addition to repeating these same circuits interleaved with a $\CZ$ at every depth in order to obtain the $\CZ$ gate fidelity estimate. In contrast, CAFE uses only 16 different circuits and 5 depths $0\le n\le8$ of $\CZ$s (with at most two additional $\CZ$ layers for state preparation and measurement).
Moreover, the single qubit gates used by IRB to create the random Clifford gates introduce unwanted error sources from decoherence and systematic errors, a problem that CAFE alleviates altogether by only repeating the cycle circuit of interest.
Note that, unlike CAFE, RB additionally provides an estimate of the average gate fidelity of all two-qubit Cliffords, which can be of independent interest.
We also want to point out that using additional depths $n$ and sampling shots can improve both approaches, and that considering a different range of noise parameters can change the resulting median absolute errors significantly. However, we have found that the conclusion remains that CAFE allows one to do a gate characterization at least as good as IRB, while requiring significantly less run time in experiment.

\section{Budgeting coherent and incoherent errors} \label{sec:coherent incoherent}
\begin{figure}
    \centering
    \includegraphics[width=0.95\columnwidth]{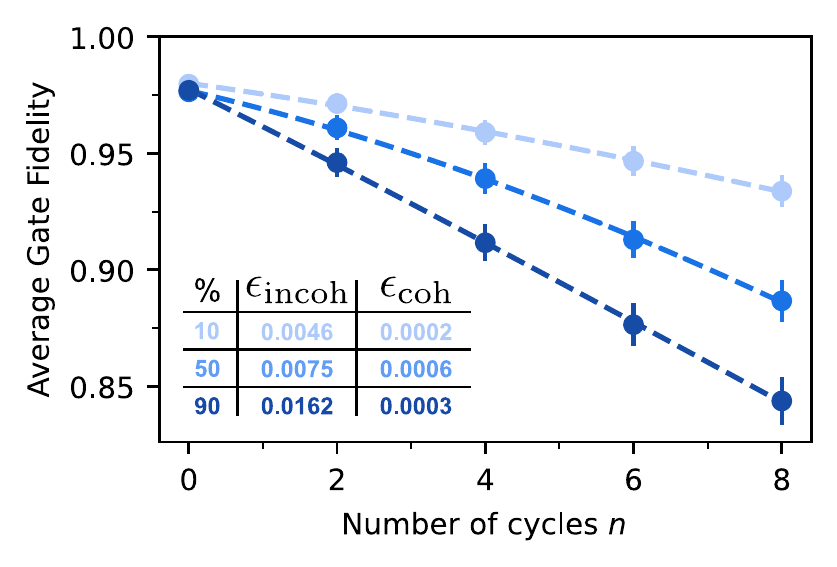}
    \caption{Experimental characterization of $\CZ$ gates performed in parallel using CAFE.
    Data represents the cycle circuit fidelity compared to the ideal $\CZ$ unitary for different cycle repetitions, along with fits using the model presented in Eq.~\ref{eq: analytical fit} and resulting gate budgets in the inset table.
    We plot the data for three specific gates -- at the 10th, 50th and 90th percentile on a Sycamore device in terms of average gate infidelity -- to illustrate the accuracy of the modeling over a wide range of data.
    Error bars represent the standard deviation of binomial distributions scaled by a factor of 5 to be visible. 
    }
    \label{fig:fitting-demo}
\end{figure}
Using CAFE, one can accurately estimate the fidelity of any unitary operation and report this single number as a performance metric, which is useful for validation and for estimating the performance of different quantum algorithms~\cite{eisert2020qcvv}.
However, such a metric alone provides very little information as to how to improve the gate fidelity in practice.
A central feature of CAFE is that one can extract actionable information about the origins of gate error by budgeting the coherent and incoherent contributions.
To do so, one can fit the fidelity decay curve to a physical model, or modify the cycle circuit to echo out different parts of the gate errors, for example by leveraging Dynamical Decoupling (DD) pulses.

We note that such budgeting is helpful for directing research focus, as interventions for coherent and incoherent errors tend to be different.

\subsection{Separating coherent and incoherent errors through fitting to a model}\label{sec:fitting model}
The CAFE experiment and resulting error budget is valid for any $m$-qubit unitary, but we will again focus on the two-qubit $\CZ$ case to simplify the discussion in the main text. 
We present a more general derivation in App.~\ref{sec:app fit}.
We assume that the cycle unitary is an excitation-preserving two-qubit gate close to a $\CZ$
\begin{align}
&\hspace{-0.7em}\tilde{U}(\Delta\theta, \Delta\gamma, \Delta\phi) = \nonumber\\[3pt]
&\hspace{-0.7em}
\scalebox{0.97}{$\begin{pmatrix}
1    &0    &0   &0\\[3pt]
0 & e^{-i\Delta\gamma}\cos (\Delta\theta) &  -ie^{-i\Delta\gamma}\sin (\Delta\theta) & 0\\[3pt]
0 &-ie^{-i\Delta\gamma}\sin (\Delta\theta) & e^{-i\Delta\gamma}\cos (\Delta\theta)& 0\\[3pt]
0 &0 &0 & -e^{-i(\Delta\phi+2\Delta\gamma)}
\end{pmatrix}$}\,,
\end{align}
where $\tilde{U}(0, 0, 0) = \CZ$, and the swap, single-qubit phase, and controlled-phased miscalibration angles are assumed to be small, such that $\Delta\theta,\, \Delta\gamma,\, \Delta\phi \ll 1$.
To simplify further the expressions here, we also assume that the noisy quantum channel implementing the cycle circuit can be described by a two-qubit depolarizing channel
\begin{align}\label{eq:depol channel}
    \mathcal{E} (\rho) = &(1 - \pdepol)\, \tilde U \rho\, \tilde U^\dag \,+\, \pdepol\, I_\dd/\dd\,,
\end{align}
which outputs a totally mixed state with probability $\pdepol$ and otherwise applies the cycle unitary $\Tilde{U}$.
With such a channel, the average gate fidelity for $n$ repetitions of the cycle circuit is given by
\begin{align}
    \begin{multlined}[t]
    \mathcal{F}_n = \frac{1}{4} - \espam - \frac{1}{20} (1-\pdepol)^n \,\cdot\\
    \Big(1 - \left|1+2e^{-i\nn\Delta\gamma}\cos(\nn \Delta\theta)+e^{-i \nn (2\Delta\gamma + \Delta\phi)} \right|^2 \Big),
    \label{eq: analytical fit}
\end{multlined}
\end{align}
where we have explicitly included the SPAM errors $\espam$, which are assumed to vary slowly over the timescale of an experiment. We note that in this model we made a steady-state assumption about the gates in our system, but modeling transient behavior could be achieved with a more advanced fit.
Using the expression in Eq.~\ref{eq: analytical fit}, we can model the experimental data to obtain the gate fidelity $\mathcal{F}$, in addition to the incoherent errors $\eincoh$ and coherent errors $\ecoh$ of the quantum operation.
To get these parameters in a way that is robust to SPAM errors, we use 
\begin{align}
    1-\mathcal{F} &= 1-\frac{\mathcal{F}_1}{(1-\espam)},\label{eq: budget fidelity}\\
    \eincoh &= 1-\frac{\mathcal{F}_1(\pdepol=0)}{(1-\espam)},\label{eq: budget incoh}\\
    \ecoh &= 1-\frac{\mathcal{F}_1(\dtheta=\dgamma=\dphi=0)}{(1-\espam)}\label{eq: budget coh}.
\end{align}
In numerical simulations, which are presented in App.~\ref{sec:app simulations}, this analysis procedure was shown to be valid and robust to different unitary errors, as well as amplitude and phase damping channels.
We use this method to budget errors in~\cref{fig:app simu irb cafe,fig:fitting-demo,fig:cafe-DD,fig:FDD-vs-XEB}.
A similar budgeting approach for single-qubit $\X(\pi)$ gates, together with experimental results acquired on a Sycamore chip, are presented in App.~\ref{sec:app 1q cafe}.

As shown in Fig.~\ref{fig:fitting-demo}, the simple model of Eq.~\ref{eq: analytical fit} allows us to accurately fit the CAFE curves spanning a wide range of $\CZ$ gate fidelities executed in parallel on a Sycamore chip.
Additional data showing the consistency of the resulting error budget with XEB is presented in App.~\ref{sec:app data}.

A useful and intuitive picture to analyze the CAFE data is to consider the incoherent and coherent errors as linear and quadratic contributions to the gate infidelity, respectively.
This can be seen directly by expanding Eq.~\ref{eq: analytical fit} up to terms $\mathcal{O}((\Delta\theta)^4, (\Delta\gamma)^4, (\Delta\phi)^4, \pdepol^2)$
\begin{align}\label{eq: quad fit}
    \mathcal{F}_n &\approx1 - \epsilon_\mathrm{spam} - \epsilon_\mathrm{lin}\, n - \epsilon_\mathrm{quad}\, n^2\\
    \epsilon_\mathrm{lin}&=\frac{3\pdepol}{4}\; n\\
    \epsilon_\mathrm{quad} &= \frac{8[(\Delta\theta)^2+(\Delta\gamma)^2+\Delta\gamma\Delta\phi]+3(\Delta\phi)^2}{20}\; n^2
\end{align}
This approximate quadratic form can be found for different cycle unitaries under similar noise channels, and could be used directly in the budgeting procedure given low gate infidelities and shallow cycle repetitions $n$, or in cases lacking an accurate analytical model for the quantum channel, however we will stick to the model in \cref{eq: budget fidelity,eq: budget incoh,eq: budget coh} in the rest of this paper.

The quantity $\eincoh$ estimates the average gate infidelity when no coherent control errors are present.
This useful characterization metric is defined as $R(\mathcal{E})$ in Ref.~\cite{wallman2015estimating}, where the authors show that it bounds the \textit{unitarity} of the channel $u(\mathcal{E})$: 
\begin{equation}
    \frac{d-1}{d}(1 - \sqrt{u(\mathcal{E})}) \leq R(\mathcal{E}).
\end{equation}
As such, we can also relate our incoherent error estimate to unitarity.
For instance, in the case of depolarizing noise, the unitarity is
\begin{equation}
    u(\mathcal{E}_\mathrm{depol}) = (1 - \pdepol)^2,
\end{equation}
and the incoherent error obtained from CAFE is, as derived in~\cref{sec:app fit},
\begin{align}
        1-\eincoh &= \frac{1}{d} + \frac{d-1}{d} (1-\pdepol)\\[5pt]
        &= \frac{1}{d} + \frac{d-1}{d} \sqrt{u(\mathcal{E}_\mathrm{depol})}. 
\end{align}

\subsection{Isolating Coherent Channels using Dynamical Decoupling}\label{sec:decaf}
\begin{figure}
    \centering
    \includegraphics[width=0.9\columnwidth]{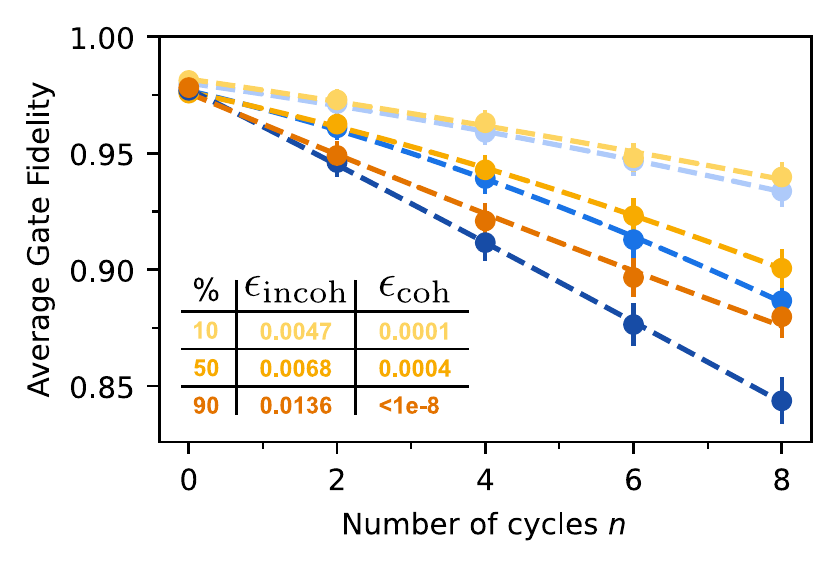}
    \caption{
    Experimental results on characterizing $\CZ$ gates in parallel when dynamical decoupling gates are part of the cycle circuit (orange), alongside the standard CAFE data presented in Fig.~\ref{fig:fitting-demo} (blue). Interleaving the $\CZ$ gates with $\X$ gates on both qubits echoes out low-frequency $Z$ noise, which contributes to $\eincoh$, and removes the sensitivity to single-qubit phase unitary errors, which contribute to $\ecoh$. Performing the simple DECAF experiment thus provides valuable information about the error mechanisms in $\CZ$ gates.
    Error bars are scaled by a factor 5 to be visible.
    }
    \label{fig:cafe-DD}
\end{figure}
Exploiting the versatility of CAFE, we can modify the cycle circuit we are characterizing in order to isolate specific error channels.
In particular, this is viable when the impact of the modifications is insignificant relative to the error channels being isolated.
As a relevant example, we can leverage the fact that the cycle circuit is repeated $n$ times by inserting dynamical decoupling gates in between repetitions.
This approach of combining DD and CAFE, which we refer to as \emph{DECAF}, is particularly useful to characterize the amount of certain coherent error present in the cycle circuit without necessitating full unitary tomography. We note that this is somewhat similar to the work in Ref.~\cite{sheldon2016characterizing}, with a single repetition of the cycle circuit, and the randomized unitaries replaced with specific DD pulses.

In the case of a $\CZ$ gate, adding an $\X$ gate to both qubits in the CAFE cycle circuit echos out the single-qubit phase errors, in addition to mitigating low-frequency noise, as demonstrated in App.~\ref{sec:app dd}~\cite{viola1998dynamical, viola2003robust}.
As shown in Fig.~\ref{fig:cafe-DD}, the DECAF data has higher fidelities and decreases more linearly in practice.  
Looking at the resulting fits over all characterized $\CZ$ gates, we see a significant decrease in the median coherent error from $5.5\times 10^{-4}$ to $9.4\times 10^{-5}$, which highlights single-qubit phase miscalibrations, alongside a decrease of the median incoherent error from $7.2\times 10^{-3}$ to $6.7\times 10^{-3}$, which indicates the level of low-frequency noise in the device.

\section{Validating unitary characterizations}\label{sec:comparing unitaries}
\begin{figure}
    \centering
    \includegraphics[width=0.9\columnwidth]{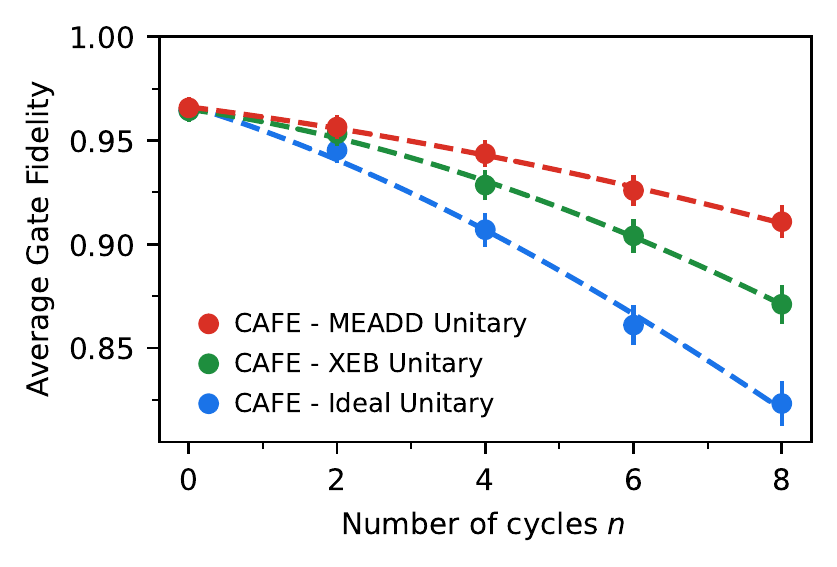}
    \caption{Using CAFE to validate different unitary characterizations of a $\CZ$ gate with 50th percentile infidelity.
    The different curves represent the fidelity between the experimental gate and an ideal $\CZ$ unitary (blue), a unitary extracted from an XEB experiment (green), and a unitary characterized with a newly introduced method called MEADD~\cite{gross2023MEADD}.
    In these three different CAFE experiments, only the final pre-measurement circuit layer changes (see the green box in Fig.~\ref{fig:schematic}).
    The improvement in fidelity can thus entirely be attributed to considering a gate unitary that is closer to the experimental $\CZ$ implementation.
    Error bars are scaled by a factor 5 to be visible.
    }
    \label{fig:FDD-vs-XEB}
 \end{figure}

One key difference between CAFE and other methods for extracting error rates is the final unentangling step before measurement.
By doing all of the inversion in a single step, similar to RB, but using the reference unitary for the inversion, we can validate different unitary characterizations, while respecting the non-Clifford nature of most coherent error models.
As such, we can use CAFE to benchmark unitary characterizations, simply by changing the final measurement step and seeing which predictions most accurately map the final state back to $|0\rangle^{\otimes m}$, in conjunction with the methods presented in Sec.~\ref{sec:fitting model}.

In Fig.~\ref{fig:FDD-vs-XEB}, we compare the CZ unitary extracted by XEB to the one extracted using a unitary characterization method called Matrix-Element Amplification by Dynamical Decoupling (MEADD)~\cite{gross2023MEADD}, which is a characterization technique that we have developed based on Floquet characterization~\cite{arute_observation_2020, neill_accurately_2021}. MEADD allows us to isolate and precisely measure the different unitary parameters in any phased fSim gate (a general excitation number preserving two-qubit gate).
We can see very clearly that the unitary predicted by MEADD is significantly better at predicting the coherent error than the unitary extracted by XEB. 
By increasing the amount of context around the gate, for example including some microwave operations or measurements on the surrounding qubits to include crosstalk or measurement-induced dephasing effects, we can see which characterizations break down in other contexts (see App.~\ref{app: surface code}), and build trust that the structures seen in our characterizations are the dominant effects on the processor impacting algorithm performance.

\section{Conclusion}\label{sec:conclusion}
CAFE separates itself from other gate characterization methods due to its simplicity and flexibility, most notably as a complementary tool to other gate characterizations that provide more granular output. We have found its ability to split coherent and incoherent errors more reliable than other methods, and the ability to experimentally test unitary characterizations has allowed us to design and evaluate novel characterization methods more effectively.

Its flexibility allows for many other as-of-yet unexplored variations, from creating fits which include the impact of leakage, to changing the cycle circuit round-by-round to echo out components in different ways, to analyzing the different input bases separately to extract details about the error structure. Our hope is that other researchers will be able to create their own modifications of the CAFE framework to study the errors facing their own systems.

\section{Acknowledgements}
We are grateful to the Google Quantum AI team for building, operating, and maintaining software and hardware infrastructure used in this work. The authors would like to thank Will Livingston, Sergio Boixo, Ben Chiaro, Vinicius S. Ferreira, Abraham Asfaw, and Paul V. Klimov for discussions and feedback on the draft, and Catherine Erickson and Andreas Bengtsson for software support. 
% The Sycamore chip used to run the experiments in this paper was designed, fabricated, assembled, and maintained by the Google Quantum AI team. 

% \clearpage
\bibliographystyle{apsrev4-1_with_title}
\bibliography{References}
\onecolumngrid
\appendix
\clearpage

\section{Preparing and measuring two-qubit states}\label{sec:2q prep}
As an aside, we describe a method to prepare or measure an arbitrary two-qubit state with a single maximally entangling operation. We start by considering a general two-qubit target state 
\begin{equation}
    \begin{split}
        |\psi\rangle = &A |00\rangle + B |01\rangle + C |10\rangle + D |11\rangle,\\
        &|A|^2 + |B|^2 + |C|^2 + |D|^2 = 1.
    \end{split}
\end{equation}
We can then write a matrix with the amplitudes of this state
\begin{equation}
    M_\psi = 
    \begin{bmatrix}
        A & B\\
        C & D
    \end{bmatrix},
\end{equation}
and perform the singular value decomposition (SVD) $M_\psi = U_\psi S_\psi V^\dagger_\psi$.
Considering the initial singular value, $S_{\psi}^{00}$, we can quantify the level of entanglement in the targeted state.
The first step to generate this state is to prepare a state with a matching entanglement signature.

\begin{figure}[h!]
    \centering
    \includegraphics[width=0.3\linewidth]{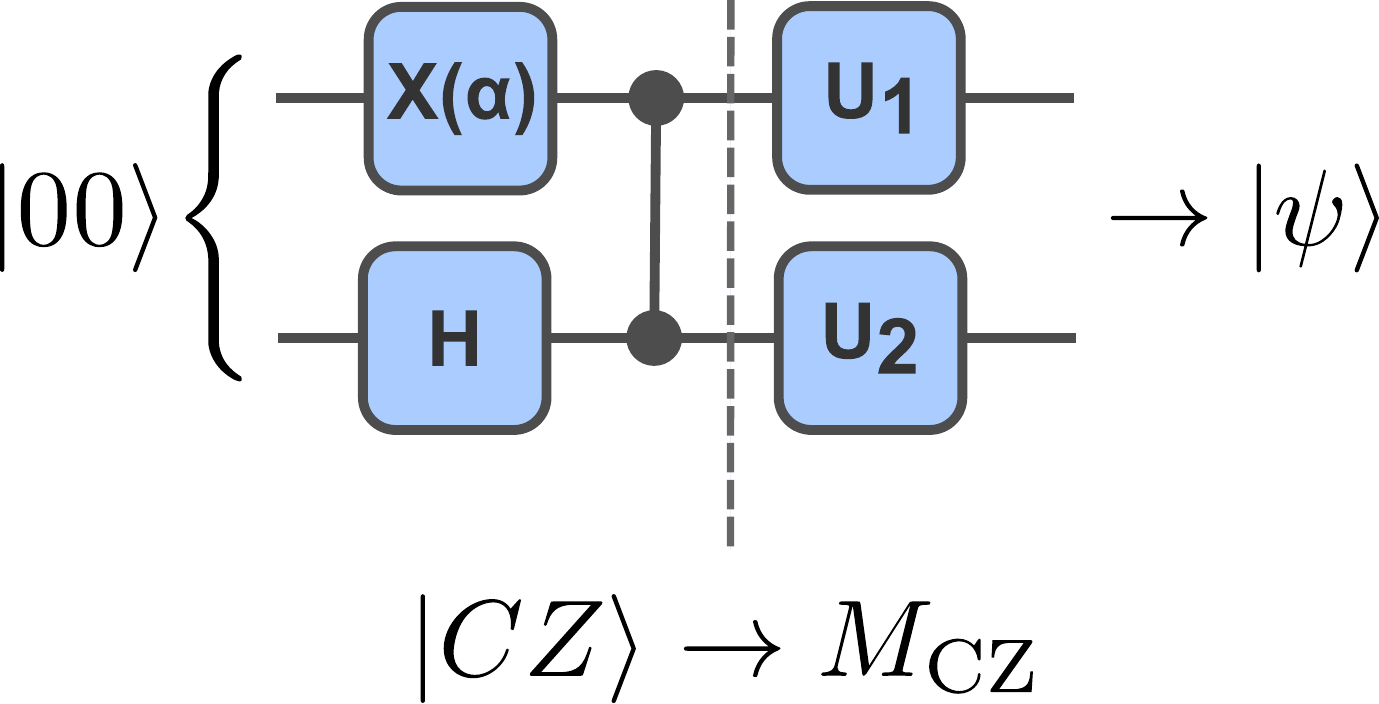}
    \caption{A simple circuit that maps $|00\rangle$ to an arbitrary two-qubit state $|\psi\rangle$ using only one $\CZ$ operation.
    The overlap of any two-qubit state with the state $|\psi\rangle$ can be obtained by executing the inverse of this circuit and reporting the probability of measuring $|00\rangle$ afterwards.
    }
    \label{fig:2q state measurement gadget}
\end{figure}

In the case where we intend to use a $\CZ$ gate, this can be done by putting one qubit in $|+\rangle$, and applying a $Y$ rotation to the other qubit with an angle of $\alpha = 2 \arccos \left(S_{\psi}^{00}\right)$.
This state, labeled as $|\CZ\rangle$ in Fig.~\ref{fig:2q state measurement gadget}, is equivalent to the final state up to single qubit rotations. 
To find these rotations, we take the SVD of the $2\times 2$ matrix corresponding to this intermediate state, $U_{CZ}S_{CZ}V^\dagger_{CZ}$.
The single qubit unitary required for the first qubit is given by $U_1 = U_\psi U_{CZ}^{-1}$, and the unitary for the second qubit is $U_2 = V_\psi V_{CZ}^{-1}$. The resulting circuit is shown in Fig.~\ref{fig:2q state measurement gadget}.
Constructions for $\CZ$ and $\sqrt{\iSWAP}$ are implemented as the methods \texttt{prepare\_two\_qubit\_state\_with\_cz} and \texttt{prepare\_two\_qubit\_state\_with\_iswap} in the open-source software \texttt{Cirq}~\cite{cirq_developers_2022_6599601}.

\section{Derivation of Eq.~\ref{eq: analytical fit}}\label{sec:app fit}
In this section, we derive Eq.~\ref{eq: analytical fit} as an example of how one could do the same for other unitaries and error models of interest.
Given a general quantum channel described by a set of Kraus operators $K_\alpha$,
\begin{align}
\mathcal{E}(\rho) = \sum_\alpha K_\alpha\rho K_\alpha^\dag\,,
\end{align}
the average fidelity of the quantum operation $\mathcal{E}$ with respect to a unitary operation $U$ is
\begin{align}
\mathcal{F}(\mathcal{E}, U) &= \int \bra{\psi(\bm\xx)}U^\dag \mathcal{E}\bigl(\, \proj{\psi(\bm\xx)}\,\bigr) U\ket{\psi(\bm\xx)}\,\dif \mu(\bm\xx)\\
&= \sum_\alpha \int \big\lvert\bra{\psi(\bm\xx)} U^\dag \KK_\alpha \ket{\psi(\bm\xx)}\big\rvert^2\,\dif \mu(\bm\xx)\,,
\end{align}
where $\bm\xx$ is a parametrization of pure quantum states and $\mu(\bm\xx)$ is the Haar measure.  
To evaluate $\mathcal{F}$, we introduce the projector on the symmetric subspace of the system and its replica, which is expressed as
\begin{align}
P_S &= \frac{\dd(\dd+1)}{2}\! \int \proj{\psi(\bm\xx)}\otimes \proj{\psi(\bm\xx)}\,\dif \mu(\bm\xx)
\,,
\end{align}
where $\dd=2^m$ is the dimension of the $m$-qubit Hilbert space.
Using this projector, the fidelity takes the form 
\begin{align}
\mathcal{F}(\mathcal{E}, U) &= \frac{2}{\dd(\dd+1)} \tr\Bigl(P_S \sum_\alpha \KK^\dag_\alpha U \otimes U^\dag \KK_\alpha \, \Bigr)\\
&= \frac{1}{\dd\xssp(\dd+1)}\sum_\alpha \tr K_\alpha^\dag K_\alpha + \big\lvert\tr (U^\dag K_\alpha)\big\rvert^2 \\
&= \frac{1}{\dd+1} + \frac{1}{\dd\xssp(\dd+1)}\sum_\alpha \big\lvert\tr (U^\dag K_\alpha)\big\rvert^2\label{eq: general agf}\,.
\end{align}
In order to compute the fidelity expression in Eq.~\ref{eq: general agf}, one computes the eigenvalues of $U^\dag K_\alpha$ and sum them up directly to get $\tr (U^\dag K_\alpha)$.
For example, we consider the case where the operation of interest can be described by a quantum channel that either applies the unitary $\tilde U$, or totally depolarizes the qubits with probability $\pdepol$
\begin{align}
\mathcal{E} (\rho) = (1 - \pdepol)\, \tilde U \rho\, \tilde U^\dag + \pdepol\, I_\dd/\dd\,,
\end{align}
where $I_\dd$ is the $\dd$-dimensional identity operator.
The fidelity of this channel after $\nn$ consecutive applications is given by 
\begin{align}\label{eq:agf_n_depolarization}
\mathcal{F}_\nn &=  (1 - \pdepol)^\nn\,\frac{\dd + \big\lvert\tr [( U^\dag)^\nn \tilde U^\nn]\big\rvert^2}{\dd(\dd+1)} + \bigl[1-(1 - \pdepol)^\nn\bigr]\,\frac{1}{\dd}\,.
\end{align}
After subtracting $\epsilon_\mathrm{spam}$ from the RHS to account for experimental SPAM errors, the expression in Eq.~\ref{eq:agf_n_depolarization} can be used to fit the CAFE data of any $m$-qubit unitary $\tilde U$, subject to symmetric depolarizing noise, with respect to a reference unitary $U$.
Note that in the case where the quantum operation under characterization, $\tilde U$, corresponds exactly with the reference unitary $U$, the CAFE data should obey the following single-exponential decay 
\begin{align}\label{eq: eincoh of pd}
    \mathcal{F}_\nn &= \frac{1}{d} + \frac{d-1}{d} (1-\pdepol)^n.
\end{align}
Otherwise, when $\tilde U \neq U$, coherent errors are present and the first term of~\ref{eq:agf_n_depolarization} will introduce errors that scale quadratically with $n$ to first order.

In the case we are interested in, we want to use CAFE to characterize a two-qubit gate, where $d=4$, $U=\CZ$ and $\tilde U$ is a number-preserving $\fSim$ gate close to a $\CZ$ gate:
\begin{align}
\tilde{U}(\Delta\theta, \Delta\gamma, \Delta\phi) = 
\begin{pmatrix}
1    &0    &0   &0\\[3pt]
0 & e^{-i\Delta\gamma}\cos (\Delta\theta) &  -ie^{-i\Delta\gamma}\sin (\Delta\theta) & 0\\[3pt]
0 &-ie^{-i\Delta\gamma}\sin (\Delta\theta) & e^{-i\Delta\gamma}\cos (\Delta\theta)& 0\\[3pt]
0 &0 &0 & -e^{-i(2\Delta\gamma+\Delta\phi)}
\end{pmatrix}\,,
\end{align}
with some residual $\SWAP$-like error captured by the angle $\Delta\theta$, single-qubit phase error captured by $\Delta\gamma$, and $\CPhase$-like error captured by $\Delta\phi$, with $\Delta\theta,\,\Delta\gamma,\,\Delta\phi\ll1$.
Since $\CZ$ is diagonal, we can show that the eigenvalues of $( U^\dag)^\nn \tilde U^\nn$ are
\begin{align}
\boldsymbol{\lambda} = 
\begin{pmatrix}
1    &e^{-i\nn(\Delta\gamma-\Delta\theta)}    &e^{-i\nn(\Delta\gamma+\Delta\theta)}   &e^{-i\nn(2\Delta\gamma+\Delta\phi)}
\end{pmatrix},
\end{align}
for $n \in \mathbb{N}$. As such, $\tr [( U^\dag)^\nn \tilde U^\nn] = \operatorname{sum}(\boldsymbol{\lambda})$ and substituting into Eq.~\ref{eq:agf_n_depolarization} gives
\begin{align}
    \mathcal{F}_n &= \frac{1}{4}-(1-\pdepol)^n \left(\frac{1-\left|1+2e^{-i\nn\Delta\gamma}\cos(\nn \Delta\theta)+e^{-i \nn (2\Delta\gamma + \Delta\phi)} \right|^2}{20} \right)\\
    &= 1 - \frac{3\pdepol}{4}\nn - \frac{8(\Delta\theta)^2+8(\Delta\gamma)^2+8(\Delta\gamma\Delta\phi)+3(\Delta\phi)^2}{20} \nn^2\\
    &\hphantom{{}\mathrel{=}}
    + \mathcal{O}((\Delta\theta)^4, (\Delta\gamma)^4, (\Delta\phi)^4, \pdepol^2)
    ,
\end{align}
which is the expression in Eq.~\ref{eq: analytical fit} of the main text, after subtracting the SPAM errors $\epsilon_\mathrm{spam}$ from the RHS to account for experimental imperfections.

\section{Single-qubit CAFE}\label{sec:app 1q cafe}
\begin{figure}
    \centering
    \includegraphics[width=0.9\columnwidth]{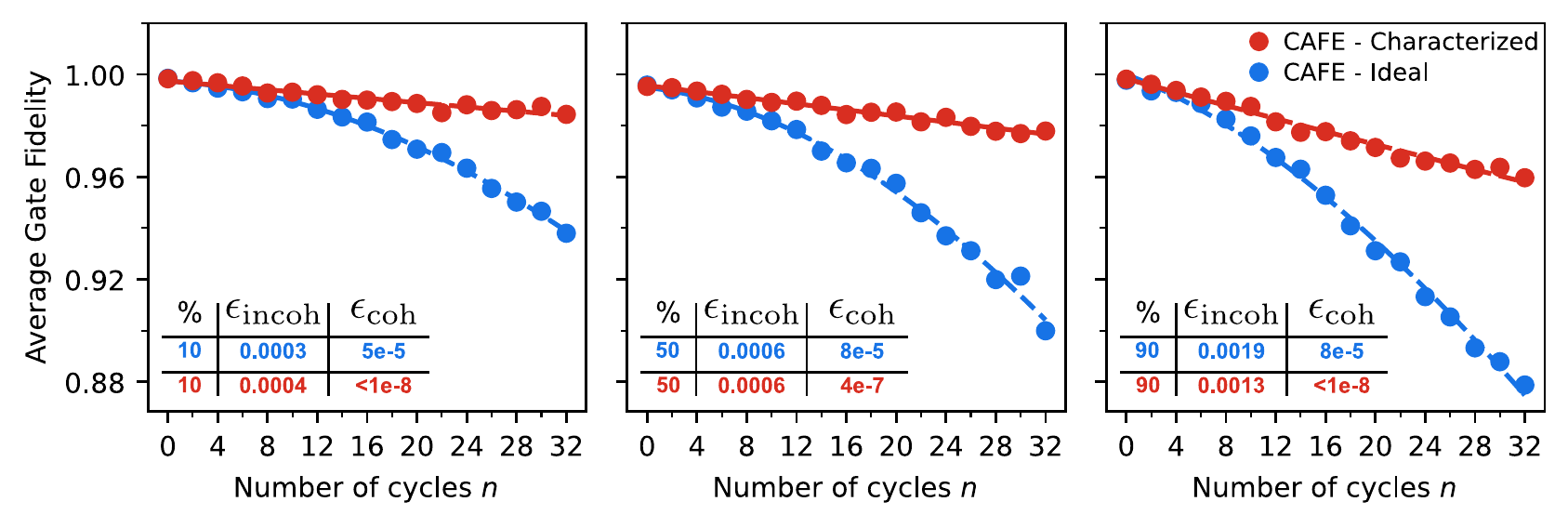}
    \caption{Single-qubit parallel CAFE results for three $\X$ gates close to the 10th (left), 50th (center), and 90th (right) percentiles of infidelity, from a total of 68 parallel $\X$ gates on a Sycamore device.
    The data in blue computes the fidelity with an ideal $U=\X$ unitary, whereas the CAFE data in red uses a characterized unitary. Note that in this data, the characterized single-qubit unitaries were not used to optimize the gate parameters, thus the imperfect calibration.
    }
    \label{fig:app sq fits}
\end{figure}
In order to give an additional example on how to deploy the CAFE characterization framework, we present an experiment characterizing 68 single-qubit $\X$ gates in parallel on a Sycamore processor.
The results are presented in~\cref{fig:app sq fits}, showing the fidelity of the experimental gate with both an ideal unitary (in blue) and a characterized one (in red), in addition to fits using an analytical model derived below, for three example gates.
In the single qubit case, the minimal 2-design contains only 4 states which can all be prepared with a single gate, which significantly speeds up the characterization.
We have used cycle repetitions $\nn$ up to $32$ here, showing how one can modify the experiment when studying especially high-fidelity operations to maintain good fit robustness.

Assuming that the incoherent noise of the experimental gate can be described by a totally depolarizing channel, our model to fit the single-qubit CAFE data can again be derived from~\cref{eq:agf_n_depolarization}.
For this case where $U=R_x(\pi)$ and $\dd=2$, we can parametrize a single-qubit unitary 
\begin{align}
\tilde{U}(\Delta\mu) &= 
\begin{pmatrix}
-\sin(\Delta\mu/2) &  -i\cos (\Delta\mu/2)\\[3pt]
-i\cos(\Delta\mu/2) & -\sin ((\Delta\mu/2)
\end{pmatrix}
\,,
\end{align}
where $\tilde U(0) = R_x(\pi)$.
One can easily show that the eigenvalues of $(U^\dag)^\nn \tilde U^\nn$ are
\begin{align}
\boldsymbol{\lambda}_\nn = 
    \begin{cases}
        (i^ne^{i\nn\Delta\mu/2}  &i^n e^{-i\nn\Delta\mu/2} ),\; \mathrm{for }\; n\; \mathrm{even}\\
        (e^{i\nn\Delta\mu/2}  &e^{-i\nn\Delta\mu/2} ),\; \mathrm{for }\; n\; \mathrm{odd}
    \end{cases}
\end{align}
where $n \in \mathbb{N}$. As such,
\begin{align}
    \left|\tr [( U^\dag)^\nn \tilde U^\nn]\right|^2 &= \left|\operatorname{sum}(\boldsymbol{\lambda}_\nn)\right|^2 = 4\cos^2(\nn \Delta\mu /2)
\end{align}
and Eq.~\ref{eq:agf_n_depolarization} becomes
\begin{align}
    \mathcal{F}_n &= \frac{1}{2} - \espam +(1-\pdepol)^n \left(\frac{2}{3}\cos^2(\nn \Delta\mu /2) -\frac{1}{6}\right)\label{eq:app fit sq}. 
\end{align}
where, as before, we have explicitly included the SPAM errors $\espam$ to account for experimental imperfections.
The error budget in terms of the fidelity $\mathcal{F}$, incoherent error contribution $\eincoh$, and coherent error contribution $\ecoh$ is obtained the same way as in~\cref{eq: budget fidelity,eq: budget incoh,eq: budget coh} of the main text, i.e. by fitting the data with this expression, evaluating it at $n=1$ with different noise parameters set to zero, and normalizing by the SPAM errors.
\Cref{fig:app sq fits} shows that such a model allows us to accurately reproduce the CAFE data obtained in the experiment.

\section{Characterizing multi-layer circuits using CAFE}\label{app: surface code}
\begin{figure}
    \centering
    \includegraphics[width=0.95\linewidth]{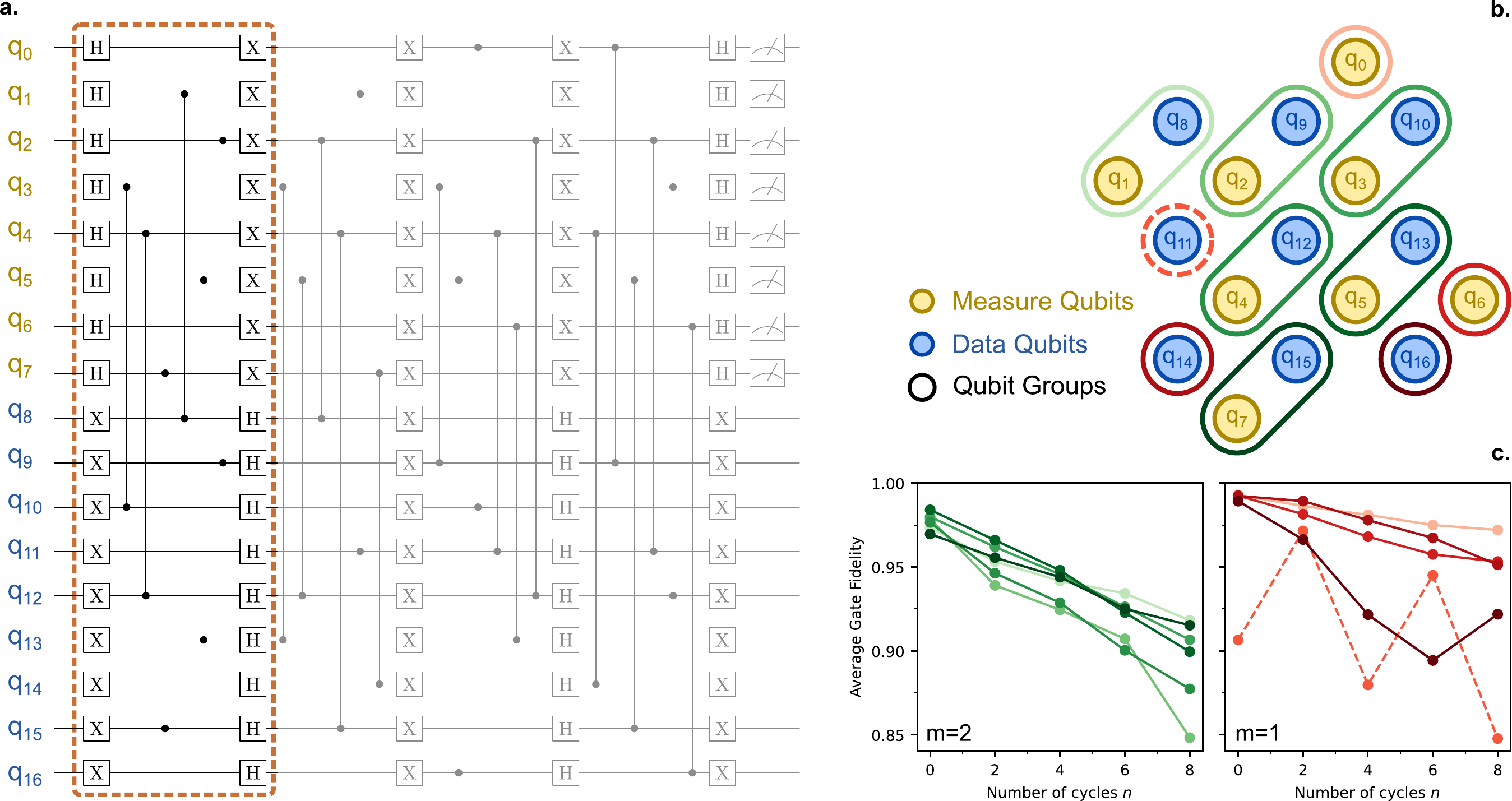}
    \caption{\textbf{a.} The stabilizer extraction circuit for a distance-3 surface code, with the section to be characterized boxed in orange. \textbf{b.} A layout for the CAFE experiment along with the qubit groups characterized. \textbf{c.} CAFE data for repetitions of the cycle circuit from \textbf{a.} on the qubit groups in \textbf{b.}, with the $m=2$ groups in green and the $m=1$ groups in red. CAFE allows us to see that $q_{11}$ (red, dotted) is experiencing a significant context-dependent error.}
    \label{fig:qec cafe}
\end{figure}

Here, we provide an example of a CAFE experiment that includes both spatial and temporal context by characterizing a cycle circuit containing multiple layers.
As illustrated in~\cref{fig:qec cafe}a, we consider a portion of the stabilizer extraction circuit for a distance-3 surface code experiment used in Ref.~\cite{https://doi.org/10.48550/arxiv.2207.06431}.
In~\cref{fig:qec cafe}b, we show how this section of the circuit can be considered on its own and broken down into qubit groups.
We can then run a parallel CAFE experiment on these qubit groups which uses the highlighted section as the cycle circuit. The resulting experimental data is shown in Fig.~\ref{fig:qec cafe}c. We can see that one of the $m=1$ qubit groups experiences significant error, despite its individual single-qubit gates $\X$ and $\Hadamard$ showing good fidelities.
This highlights the importance of context-aware characterization, as the error mode being experienced was only visible when the gates were run in conjunction. This experiment also highlights the flexibility of CAFE in being able to characterize multi-operation circuits simultaneously on qubit groups of varying size.

\section{Numerical CAFE simulations}\label{sec:app simulations}
In this section, we present numerical simulations where we characterize a $\CZ$ gate using the CAFE approach presented in the main text.
We simulate the same $2^4=16$ circuits executed on the Sycamore device with noisy two-qubit unitaries, together with either depolarizing noise in App.~\ref{sec:app-depol-simus} or amplitude and phase damping noise in App.~\ref{sec:app-t1t2-simus}.
We also used the same low cycle repetitions $n \in [0,2,4,6,8]$ which make the CAFE experiment have especially low execution time relative to other two-qubit benchmarking techniques.
In~\cref{sec:app-IRB-CAFE-simus} we showed that using CAFE to characterize a $\CZ$ gate can allow for a more accurate fidelity estimation than the widely used Randomized Benchmarking (RB) protocol while using significantly less experimental resources.
Note that we also simulate the sampling of $2000$ shots used experimentally, which places an upper bound on the standard deviation of all the CAFE data points presented in Figs.~\ref{fig:fitting-demo}, \ref{fig:cafe-DD}, and \ref{fig:FDD-vs-XEB} of
\begin{align}
    \sigma_\mathcal{F} \le \frac{1}{8\sqrt{N_\mathrm{shots}}} \approx 0.0028,
\end{align}
which is smaller than all the markers used in the plots of the main text.

Importantly, these simulations allow us to confirm the validity of the CAFE experiment and following fitting procedure to obtain accurate estimates of the following: the average gate fidelity of the quantum operation with regards to any reference unitary, the incoherent error contribution to the infidelity and the coherent error contribution.

The simulations were performed using the open-source software \texttt{Cirq}~\cite{cirq_developers_2022_6599601}.

\subsection{Depolarizing noise}\label{sec:app-depol-simus}
\begin{figure}[t]
    \centering
    \includegraphics[width=0.9\columnwidth]{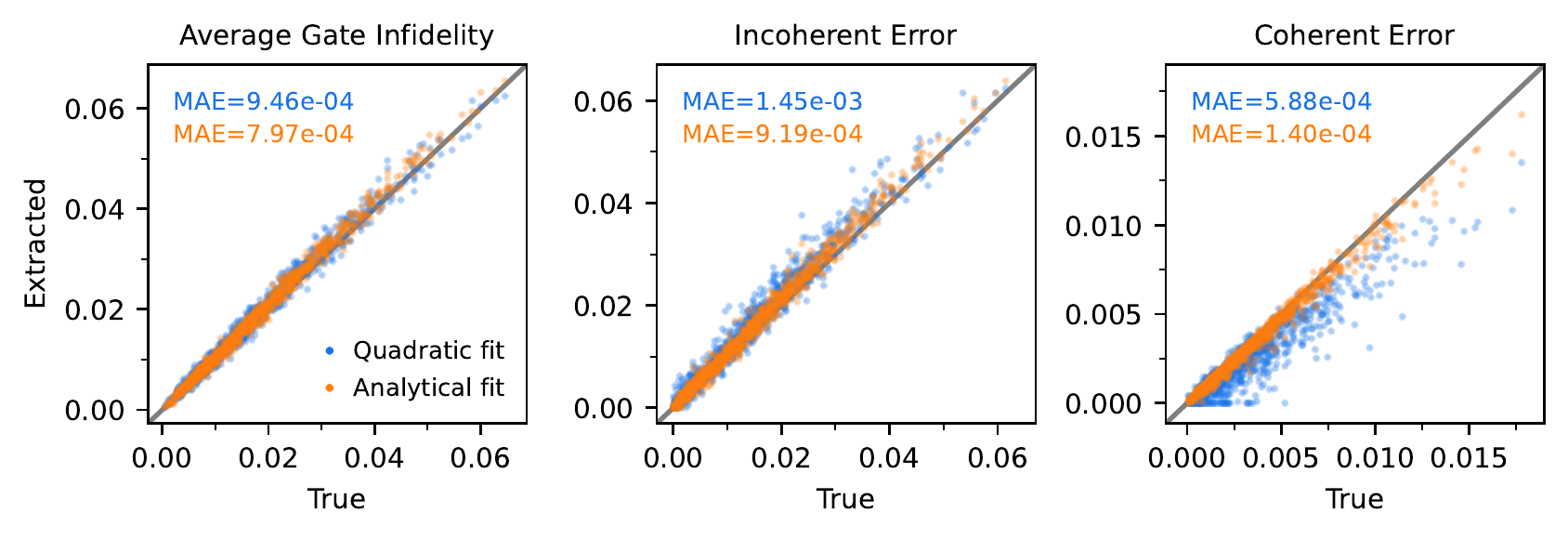}
    \caption{Simulating the CAFE experiment for characterizing 1000 different $\CZ$ gates with realistically small coherent errors and depolarizing incoherent noise.
    Two error budgeting techniques obtained by fitting this CAFE data are presented.
    In blue, using a quadratic fit for the fidelity at depths $\nn \in [0,2,4]$, and in orange using the analytical expression of~\cref{eq: analytical fit} for depths $\nn \in [0,2,4,6,8]$.
    Inset shows the median absolute errors (MAE) of the two error budgeting techniques from the true values inserted into the simulation. We note that MAE scales with the true error of the gates being considered. The true incoherent (coherent) errors are obtained from the channel's average gate infidelity when only depolarizing errors (unitary errors) are present in the simulation. 
    }
    \label{fig:app-simu-depol}
\end{figure}
In a first case, we consider that the noisy $\CZ$ gate we are trying to characterize is described by the quantum channel
\begin{align}
    \mathcal{E} (\rho) = &(1 - \pdepol)\, \tilde \VV \rho\, \tilde \VV^\dag \,+\, \pdepol\, I_\dd/\dd\,,
\end{align}
which outputs a totally depolarized state with probability $\pdepol$, and otherwise applies the excitation-preserving unitary
\begin{align}
\tilde\VV(\theta, \zeta, \chi, \gamma, \phi) =
\begin{pmatrix}
1    &0    &0   &0\\[3pt]
0 &e^{-i(\gamma + \zeta)} \cos\theta &  -i\, e^{-i(\gamma-\chi)}\sin \theta & 0\\[3pt]
0 &-i\, e^{-i(\gamma+\chi)}\sin \theta &e^{-i(\gamma - \zeta)} \cos\theta& 0\\[3pt]
0 &0 &0 &-e^{-i(2\gamma + \phi)}
\end{pmatrix}\,,
\end{align}
where we allow for miscalibrations in all of the five angles that parametrize this gate, where $\tilde\VV(0,0,0,0,0)=\CZ$.
For the simulation, we then sample the depolarizing probability uniformly in the range from 0 to 0.05, and sample the miscalibrated angles from a normal distribution with zero mean and standard deviation of 0.05~rad such that
\begin{align}
    \pdepol &\propto \mathcal{U}(0, 0.05)\\
    \{\theta, \zeta, \chi, \gamma, \phi\} &\propto \mathcal{N}(0, 0.05^2).
\end{align}
Using this model, we can simulate the noisy CAFE experiment for different cycle repetitions $\nn$, and then fit these data points to obtain an error budgeting of the noisy $\CZ$ gate, which we can compare directly with the true parameter values that were drawn for the simulation.

In Fig.~\ref{fig:app-simu-depol}, we present the average gate infidelity $(1-\mathcal{F})$, incoherent error contribution $\eincoh$, and coherent error contribution $\ecoh$ to this infidelity for $1000$ independently sampled unitaries and depolarizing probabilities.
We present two valid gate error budgeting approaches, one which fits the CAFE data with the analytical expression in Eq.~\ref{eq: analytical fit} (orange points), and one which uses the simple quadratic form of Eq.~\ref{eq: quad fit} (blue points).
These scatter demonstrate the validity of CAFE to characterize the fidelity of a quantum operation and budget its coherent and incoherent contributions with an imprecision smaller or equal to 0.001 on median for realistic gate fidelities.
Note that the reported median absolute errors depend on the specific range of gate fidelities considered. For example, the MAE values decrease significantly when considering only gate infidelities $<1\%$ in the ensemble.

The quadratic fit results shown in blue in~\cref{fig:app-simu-depol} are an indication of how the CAFE framework can be deployed in order to estimate the fidelity of an operation in context.
However, this simplest approach gives slightly biased results for $\eincoh$ and $\ecoh$, which is understood from the breakdown of the $n\pdepol\ll1$ approximation and/or the $n\alpha\ll1$ approximation, where $\alpha$ here stands for any of the five miscalibrated $\CZ$ angles.
Consequently, using a quadratic fit should be used carefully, for operations with high fidelity ($>99$~\%) and using shallow cycle repetitions $\nn$.
In fact, for these simulations, we have found that using larger depths than $n=4$ for the quadratic fit decreased the accuracy of the resulting error budgets.
However, if the fidelity of the operation of interest is very high, for example with single-qubit gates, using a quadratic fit becomes an attractive strategy when lacking a proper analytical model of the error origins.

On the other hand, the analytical fit performs well for larger depths $\nn$ since it does not rely on any approximation. Here, we have only used the depths $n \in [0,2,4,6,8]$ to be consistent with the experiment results of the main text.
These results are an important validation for CAFE, but not necessarily surprising since we are using the same error model to simulate the experiment and to fit the resulting data.
In the next sub-section, we show that this model also holds very well for different noise models such as amplitude and phase damping, provided they cause realistic gate infidelities of a few percent or less.

\subsection{Amplitude and phase damping noise}\label{sec:app-t1t2-simus}
\begin{figure}
    \centering
    \includegraphics[width=0.9\columnwidth]{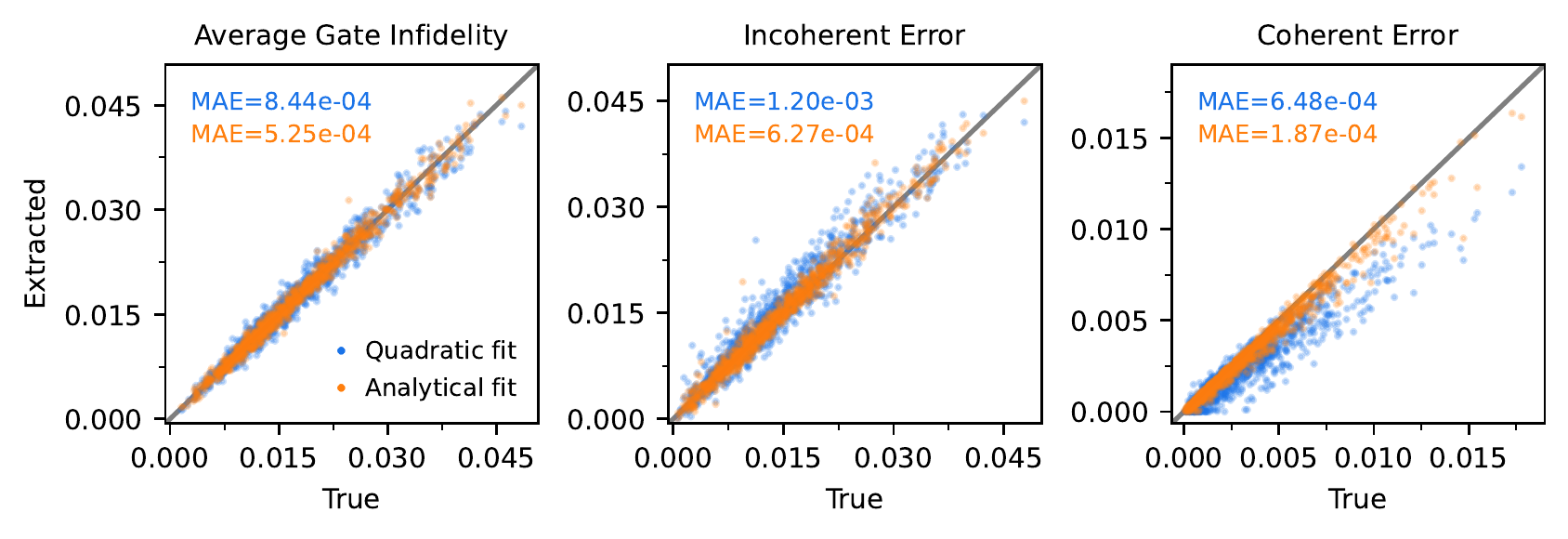}
    \caption{Similar simulations as presented in~\cref{fig:app-simu-depol}, now implementing realistic amplitude and phase damping incoherent noise instead of a depolarizing channel.
    The analytical fit, which still uses~\cref{eq: analytical fit} that effectively lumps the incoherent noise into a $\pdepol$ probability, works remarkably well even under this different noise channel.
    }
    \label{fig:app simu t1t2}
\end{figure}
To verify the robustness of the CAFE framework to different incoherent noise processes, we have also simulated a two-qubit CAFE experiment where the quantum channel is described by the same unitary $\tilde\VV$ with randomly sampled angles as before, but now followed by a single-qubit amplitude and phase damping channel on both qubits.
The total decay and phase-flip probabilities for both qubits were sampled from a normal distribution with a standard deviation of 0.03, before taking the absolute value
\begin{align}
    \{p_\mathrm{decay, total}, p_\mathrm{phase flip, total}\} &\propto \left|\mathcal{N}(0, 0.03^2)\right|,
\end{align}
and the individual decay and phase-flip probabilities of the two qubits were splitting these total probabilities with a ratio drawn from a normal distribution with mean of 0.5 and standard deviation of 0.1, such that the qubits have similar but distinct coherence properties.

In~\cref{fig:app simu t1t2}, we show that the CAFE framework we developed again gives very accurate $\CZ$ characterizations and error budgets.
In particular, using the analytical expression of~\cref{eq: analytical fit} to fit this CAFE data works remarkably well given that the noise present in the gate cannot be fully described by the assumed depolarizing channel.
Since the incoherent noise is small, but in the realistic range of creating about 0.1~\% to 4~\% gate infidelity, the fit is able to approximate well the incoherent error contribution into the single exponential of the model.
This gives us confidence that we can leverage this model in realistic gate characterization scenarios.
Moreover, the simulations show that the accuracy of using this model increases with increasing gate fidelities, which is a great indication for the continued usefulness of this method as gates and qubit coherences keep improving.

\subsection{Fitting CAFE data}
\begin{figure}
    \centering
    \includegraphics[width=0.9\columnwidth]{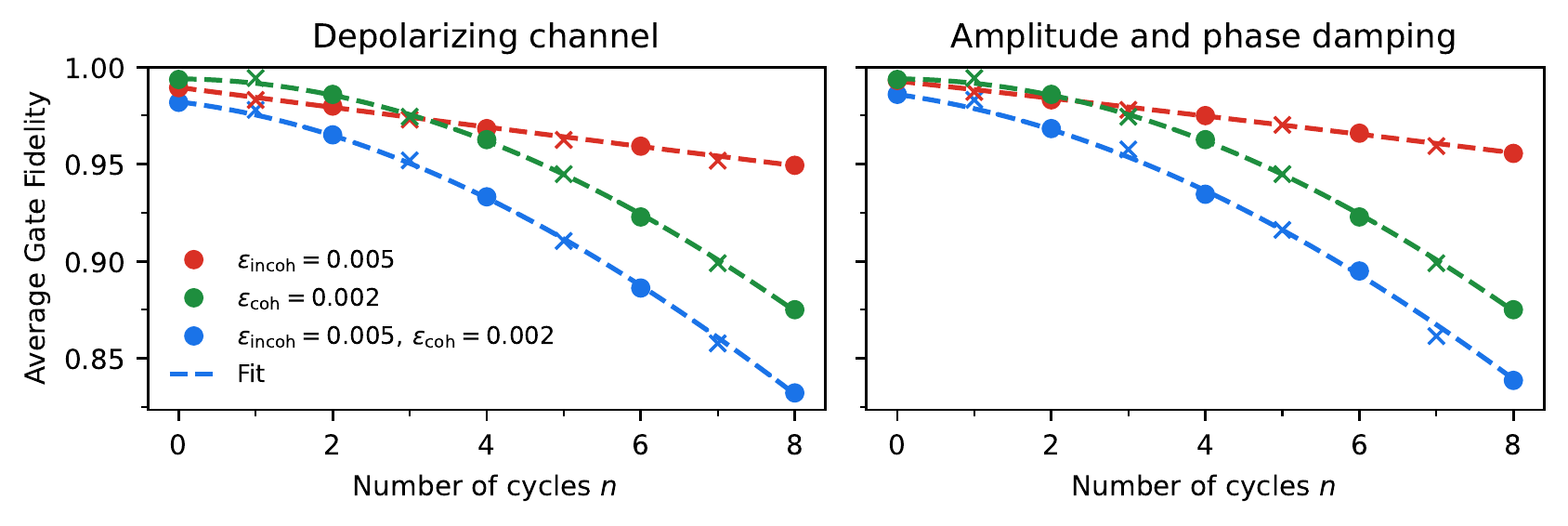}
    \caption{Example of fitting the CAFE curve obtained in simulations with the depolarizing noise channel used in~\cref{sec:app-depol-simus} (left) and the amplitude and phase damping noise channel used in~\cref{sec:app-t1t2-simus} (right).
    In blue, we show simulations including both incoherent errors of strength $\eincoh=0.005$ and coherent errors of $\ecoh=0.002$, and the dotted line shows the fit of this data using~\cref{eq: analytical fit}.
    As in the main text, only the even depths are used in the fit, the odd depths points are presented with x symbols simply to show their consistency with the model.
    Note: the green data is the same in both panels as we fix the incoherent errors to be zero.
    }
    \label{fig:app simu fits}
\end{figure}
In~\cref{fig:app simu fits}, we present examples of the CAFE data together with fits to the model of~\cref{eq: analytical fit} realized on the simulation data to obtain the $\CZ$ error budgets presented in~\cref{fig:app-simu-depol,fig:app simu t1t2}.
We see that the model of~\cref{eq: analytical fit} fits the CAFE data very well for realistic noise of strength $\eincoh=0.005$ and $\ecoh=0.002$, even when the incoherent noise comes from an amplitude and phase damping channel.
In~\cref{fig:app simu fits}, we also show simulation data using the same noise channel but an ideal unitary ($\ecoh=0$) in red, and using the same unitary without any incoherent noise ($\eincoh=0$) in green.
The linear and purely quadratic behaviors of these curves, respectively, can be easily understood from the quadratic form of~\cref{eq: quad fit} where the incoherent errors build up linearly with $n$, whereas the coherent errors build up quadratically.

Since we simulate the actual 16 circuits used to obtain the average gate fidelity at different depths, we capture some depth-dependent behaviors that are due to coherent effects in the single- and two-qubit unitaries (see for example the green `x' at $n=1$ in~\cref{fig:app simu fits}, which is actually higher than the point at $n=0$).
This is not an artifact of the finite sampling, but is due to the fact that the preparation and measurement circuits both require an imperfect $\CZ$ gate, which can coherently map the state closer or further away from the desired state $|00\rangle$, depending on the specific imperfect unitaries.
Similarly, part of the incoherent noise channel can anti-commute with the $\CZ$ unitary and produce back-and-forth behaviors between the odd and even depths $n$. We have found in simulation that using only the even depths avoids this issue, while performing similarly or better in terms of accuracy in error budgeting, compared to using all the even and odd depths.

\section{Dynamical Decoupling in CAFE}\label{sec:app dd}
In this Appendix, we show how applying an $\X$ gate to both qubits after the $\CZ$ in the cycle circuit decouples the $\fSim$ unitary from both of its single-qubit phases.
First, we can show that interleaving a pair of $\fSim$ gates with parallel $\X$ gates gives a cycle unitary that also corresponds to an $\fSim$ gate.
To see this, consider what happens when conjugating an $\fSim$ unitary by parallel $\X$s:
\begin{align}
    (X\otimes X)\fSim(X\otimes X)
    &=
    \begin{pmatrix}
        0 & 0 & 0 & 1
        \\
        0 & 0 & 1 & 0
        \\
        0 & 1 & 0 & 0
        \\
        1 & 0 & 0 & 0
    \end{pmatrix}
    \begin{pmatrix}
        1 & 0 & 0 & 0
        \\
        0 & e^{-i\gamma}u_{11} & e^{-i\gamma}u_{12} & 0
        \\
        0 & e^{-i\gamma}u_{21} & e^{-i\gamma}u_{22} & 0
        \\
        0 & 0 & 0 & e^{-i(2\gamma+\phi)}
    \end{pmatrix}
    \begin{pmatrix}
        0 & 0 & 0 & 1
        \\
        0 & 0 & 1 & 0
        \\
        0 & 1 & 0 & 0
        \\
        1 & 0 & 0 & 0
    \end{pmatrix}
    \\
    &=
    \begin{pmatrix}
        e^{-i(2\gamma+\phi)} & 0 & 0 & 0
        \\
        0 & e^{-i\gamma}u_{22} & e^{-i\gamma}u_{21} & 0
        \\
        0 & e^{-i\gamma}u_{12} & e^{-i\gamma}u_{11} & 0
        \\
        0 & 0 & 0 & 1
    \end{pmatrix},
\end{align}
where we've factored the $\fSim$ unitary into phases between the particle-number subspaces and a special unitary $U$ in the single-excitation subspace
\begin{align}
    U
    &=
    \begin{pmatrix}
        u_{11} & u_{12}
        \\
        u_{21} & u_{22}
    \end{pmatrix}
    =
    \begin{pmatrix}
        e^{-i\zeta}\cos\theta & -ie^{i\chi}\sin\theta
        \\
        -ie^{-i\chi}\sin\theta & e^{i\zeta}\cos\theta
    \end{pmatrix}.
\end{align}
Following this with a second $\fSim$ gate gives
\begin{align}
    \fSim(X\otimes X)\fSim(X\otimes X)
    &=
    e^{-i2\gamma}
    \begin{pmatrix}
        e^{-i\phi} & 0 & 0 & 0
        \\
        0 & \tilde{u}_{11} & \tilde{u}_{12} & 0
        \\
        0 & \tilde{u}_{21} & \tilde{u}_{22} & 0
        \\
        0 & 0 & 0 & e^{-i\phi}
    \end{pmatrix}
    \\
    UXUX
    &=
    \tilde{U}
    =
    \begin{pmatrix}
        \tilde{u}_{11} & \tilde{u}_{12}
        \\
        \tilde{u}_{21} & \tilde{u}_{22}
    \end{pmatrix}
    =
    \begin{pmatrix}
        u_{11}u_{22}+u_{12}^2 & u_{11}(u_{21}+u_{12})
        \\
        u_{22}(u_{21}+u_{12}) & u_{11}u_{22}+u_{21}^2
    \end{pmatrix}.
\end{align}
Here the common phase $\gamma$ has been eliminated (only appearing as a global phase), and the controlled phase $\phi$ is a relative phase between the even- and odd-parity subspaces.
For the case of something close to a CZ gate (small swap angle $\theta$ and small differential phase $\zeta$), the single-excitation unitary simplifies to
\begin{align}
    U
    &=
    I-i(\theta\cos\chi X-\theta\sin\chi Y+\zeta Z)+\mathcal{O}(\zeta^2)+\mathcal{O}(\theta^2)
    \\
    XUX
    &=
    I-i(\theta\cos\chi X+\theta\sin\chi Y-\zeta Z)+\mathcal{O}(\zeta^2)+\mathcal{O}(\theta^2)
    \\
    UXUX
    &=
    I-i2\theta\cos\chi X+\mathcal{O}(\zeta^2)+\mathcal{O}(\theta^2)+\mathcal{O}(\zeta\theta)
    \,.
\end{align}
As such, the effects of the differential phase $\zeta$, to first order, are also removed by this echoing.

\section{Swap errors without phase matching}\label{sec:app phase mathching}
When running CAFE with characterized unitaries in the experiments performed in this work, we do not attempt to correct the swap errors.
This is for two reasons.
First, the swap errors are the smallest errors in the system, and tend to be overshadowed by the other sources of error.
Second, in our system, the relative phases accumulated between qubits sitting at different idle frequencies are accounted for by changing the microwave phases.
This is possible because the two-qubit gate we are using, the CZ gate, ideally commutes with such differential phases.
The effect of not ``phase matching'' (that is, removing this differential phase by shifting qubit frequencies) on swap errors is that the swap phase $\chi$ is shifted from cycle to cycle, preventing the swaps from coherently adding and further diminishing their effect relative to the other error sources.
Should one be concerned with relatively large swap angles $\theta$, one could incorporate the application-dependent value of $\chi$ in the determination of the circuit to map the predicted state back to $|00\rangle$.

\section{Additional $\CZ$ Data}\label{sec:app data}
\begin{figure}
    \centering
    \includegraphics[width=0.85\columnwidth]{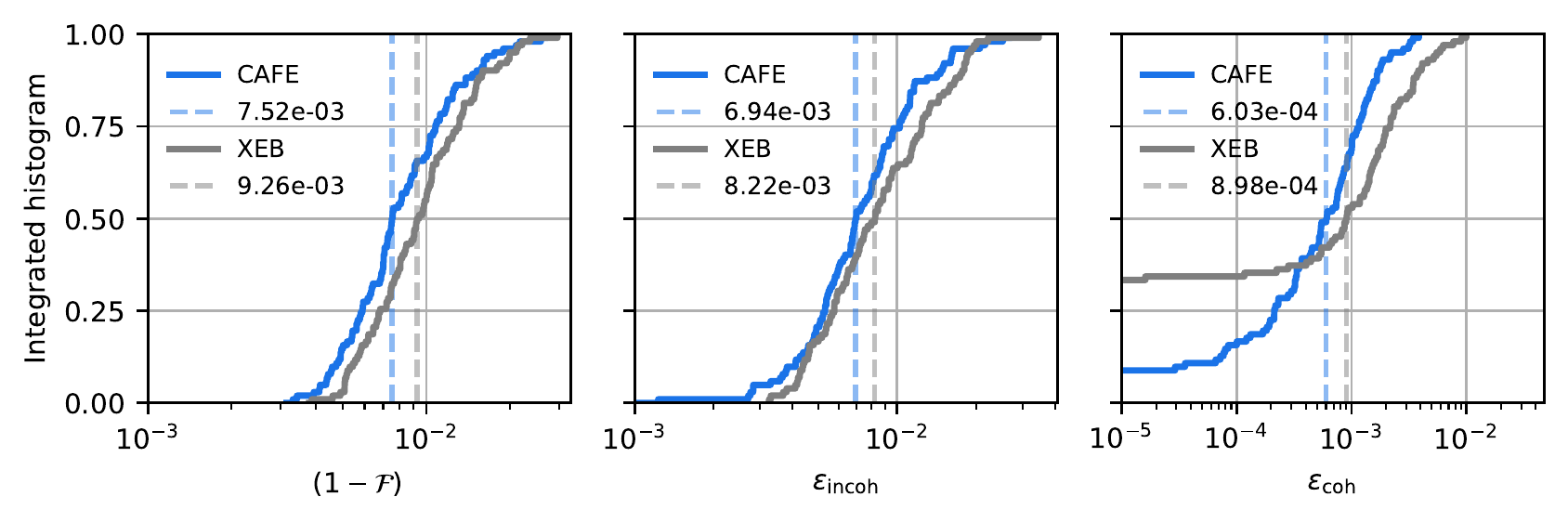}
    \caption{
    Error budgets of parallel $\CZ$ gates on a Sycamore processor.
    Data in blue is obtained from the CAFE experiment using the procedure detailed in the main text, where 3 samples of this data are shown in~\cref{fig:fitting-demo}.
    For reference, we present similar gate error budget information as obtained from standard XEB in gray, where the incoherent error is extracted from the speckle purity and the coherent error is obtained from the difference with the total gate error.
    Data shows 103 different two-qubit gates.
    }
    \label{fig:app histograms budget}
\end{figure}

For completeness, we present in~\cref{fig:app histograms budget} the entire error budget results from the parallel $\CZ$ CAFE dataset acquired on a Sycamore processor.
This data illustrates that CAFE can be straightforwardly deployed on large-scale quantum processors to characterize the fidelity of quantum operations in parallel, while budgeting the incoherent and coherent error contributions.
For reference, we also present an error budgeting of the same $\CZ$ gates obtained from XEB. Note that in this latter case, the coherent error contribution is obtained by subtracting the incoherent speckle purity error from the total XEB error, which gives an unphysical negative $\eincoh$ value for 33 out of the 103 gates. In our tests, this problem did not arise for CAFE.

%congrats! you found the secret comment. See a preview of our MM2023 talk at https://www.youtube.com/watch?v=dQw4w9WgXcQ.

\end{document}